\documentstyle[epsf,epsfig,twocolumn,prl,aps]{revtex}

\newcommand{\be}{\begin{equation}}
\newcommand{\ee}{\end{equation}}
\newcommand{\bea}{\begin{eqnarray}}
\newcommand{\eea}{\end{eqnarray}}

\newcommand{\lton}{\mathrel{\lower.9ex
                  \hbox{$\stackrel{\displaystyle <}{\sim}$}}}

\begin{document}

\title{High-Multiplicity $pA$ collisions and the small-$x$ effective action}
\author{A.\ Dumitru}
\address{
Department of Physics, Brookhaven National Laboratory, 
Upton, NY 11973}

\maketitle   

\begin{abstract}
I discuss the $p_t$ distributions for high-multiplicity events
originating from semi-classical variation of the gluon density of the proton.
The multiplicity distribution measures the curvature of
the effective action for the small-$x$ gluon fields.
For $pA$ collisions at the RHIC and LHC colliders, semi-classically the
multiplicity distribution reflects the distribution of saturation
momenta of the proton but not that of the
nucleus. The average transverse momentum in the central region grows with
$dN/dy$, while the $p_t$ distribution of leading hadrons in the proton
fragmentation region should depend less on the multiplicity in the central
region.\\
\end{abstract}

\narrowtext      

High-energy hadronic scattering requires understanding of the
non-abelian gauge fields of hadrons at small $x$, which is the fractional
light-cone energy carried by the quanta of the fields.
At very high energy, $\log 1/x\gg1$, the gluon fields in a hadron become
very strong, corresponding to high gluon density.
This is where one expects that cross sections become comparable to the
geometric size of the hadron and where the unitarity limit is reached. 
A perturbative QCD based mechanism for unitarization of cross sections is 
provided by gluon saturation effects~\cite{glr}. 
A semi-classical approach to gluon saturation and QCD at high energy
was developed in~\cite{mv,yk,balit,jk,ianc,Mueller:1999wm,KovWied}
and will be applied here to high energy proton-nucleus collisions.

Large nuclei facilitate the study of gluon saturation
effects because the gluon density per unit transverse area
is larger than in a proton. The scale associated
with the high gluon density, the saturation scale $Q_s$, grows
with energy and atomic number $A$.
At a resolution less than $Q_s^2$, the color field carries large occupation
numbers, of order of the inverse QCD coupling constant, $1/\alpha_s$. Thus,
the nuclear wave function at $Q^2<Q_s^2$ resembles a Bose ``condensate''. The
local color charge density in the transverse plane is a stochastic variable
which eventually has to be averaged over,
see below. Also, by Lorentz time dilation
the large $x$ gluons evolve slowly and so for the small-$x$
gluons they appear as a ``frozen'' source near the light cone.
Therefore, the high gluon density state of QCD at 
$Q^2\lton Q_s^2$ is called a ``Color Glass Condensate''~\cite{ianc}.

The small-$x$ gluon fields of
hadrons or nuclei can be treated semi-classically because the occupation
numbers parametrically are of order $1/\alpha_s\gg1$. One can
integrate out the fields at large $x$ whose dynamics is ``frozen'',
thereby generating an effective action for the small-$x$ 
gluons~\cite{mv,yk,balit,jk,ianc,Mueller:1999wm,KovWied},
\be\label{eff_act}
S[\rho] = \int dy\,d^2x_t \, V[\rho]~,
\ee
with $y$ being the rapidity.
The large-$x$ gluons effectively act as a source of color charge in the
Yang-Mills equations of motion for the small-$x$ fields; $\rho$ denotes the
color charge density per unit transverse area $d^2x_t$, and rapidity
$dy$. This is a stochastic variable which eventually has to be
integrated over,
\be\label{aveO}
\langle O\rangle = \int{\cal D}\rho\, \exp(-S[\rho])\, O[\rho]~,
\ee
where $O[\rho]$ denotes some observable which is a functional of $\rho$.
In their original papers~\cite{mv},
McLerran and Venugopalan suggested the simplest
possible effective potential
\be\label{quadpot}
V_2[\rho;\mu^2] = \frac{{\rm tr}\, \rho^2}{\mu^2}~.
\ee
This is the lowest-dimensional operator.
In~(\ref{quadpot}), $\mu^2$ is
a real number, and physically is simply the mean square fluctuation of the
density of ``hard'', large-$x$ gluons in the source.
It determines the curvature of the effective potential at the minimum.

\begin{figure}[htp]
\centerline{\hbox{\epsfig{figure=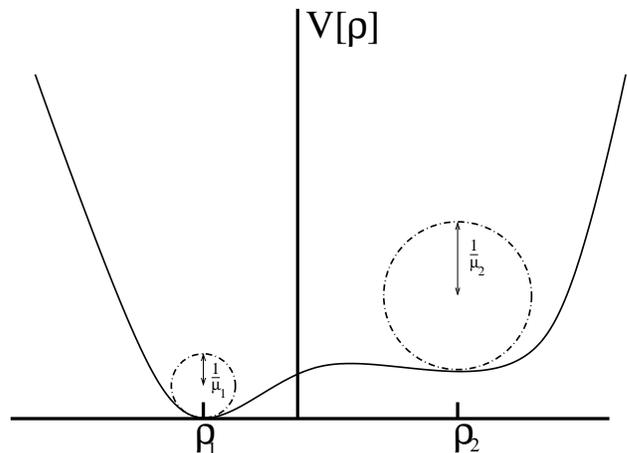,height=6cm}}}
\caption{Double-well potential for the color charge density per unit area.}
  \label{fig_DW}
\end{figure}
In principle, one might try other choices for the effective potential
$V[\rho]$, c.f.~\cite{Pirner:2002fe,mahlon}, as for example
the double well potential depicted in Fig.~\ref{fig_DW}, which looks just like
the effective potential for a first-order phase transition in thermal
$SU(3)$ gauge theory near $T_c$~\cite{finiteT}.
The curvature of the potential in the first well at $\rho_1$ is $1/\mu_1$,
and that in the second well is $1/\mu_2$.
Semiclassically, one considers only small fluctuations about the states which
minimize the action. Thus, for this potential the averaging over $\rho$
from eq.~(\ref{aveO}) turns into\footnote{The operator $O$ is assumed to be
defined such that tadpole contributions are subtracted; for example, for
the two-point function $\langle\rho(x_t,y)\rho(z_t,y')\rangle-\rho_i^2$.
One can then shift variables in~(\ref{ave_DW}), $\rho\to\rho+\rho_i$.}
\be\label{ave_DW}
\sum\limits_{i=1,2} w_i \int{\cal D}\rho \, \exp\left(
-\int dy\,d^2x_t \, V_2[\rho-\rho_i;\mu_i^2]\right) \, O[\rho]~.
\ee
The probabilities for the
configurations $\rho_1$ or $\rho_2$ are determined by the classical action:
\be
w_i = \frac{\exp(-S[\rho_i])}{\exp(-S[\rho_1])+\exp(-S[\rho_2])}~.
\ee
Note that if the effective potential is quadratic for small fluctuations
about each of its minima, renormalization-group evolution presumably
preserves that functional form~\cite{yk,balit,jk,ianc}. That is, imposing
the potential from Fig.~\ref{fig_DW} as the initial condition (at small
rapidity), it will keep its shape at larger rapidity, and the
ratio of the curvatures at the two minima stays the same.

One can evidently generalize the above to effective potentials with several
classical minima. One could also imagine that there is a continuum of nearly
degenerate states near a classical minimum. For example, 
at a tricritical point the lowest-dimensional operator
appearing in the action is $\sim{\rm tr}\rho^6$. The state of lowest action is
that for $\rho=0$ but the potential about that classical state is flat, and so
states in its vicinity are nearly degenerate.
This type of potential might not be preserved by the RG evolution in rapidity,
though.

The distribution $w(\mu^2)$ of correlation lengths determines the
multiplicity distribution in hadronic collisions. Here, I do not attempt to
compute that distribution. Rather, given a nonvanishing probability for
a class of events with higher multiplicity $dN/dy$ than the
average, I discuss the origin of such events and how the $p_t$ distribution
in that class of events is different from the average. Nevertheless, it
will be very interesting of course to analyze the multiplicity distribution
experimentally; for example, at a tricritical point events
with larger than average multiplicity could be due to semi-classical
fluctuations of the color charge density.
A potential with several discrete local minima as in Fig.~\ref{fig_DW}
would lead to multiple peaks in the multiplicity distribution, unless the
curvature radii of the potential are the same. For finite temperatures,
the effect of a second local minimum in the effective
potential on multiplicity and momentum distributions of produced particles
has been discussed previously using the hydrodynamical model~\cite{mult}.

Henceforth, I consider
``$pA$'' collisions, i.e.\ collisions of a dilute projectile with a dense
target. For simplicity, $\mu^2$ for the nucleus is chosen to 
be constant over the transverse plane. For large nuclei ($A\simeq200$)
this should not be a bad approximation. Realistic transverse density
distributions can be treated numerically~\cite{KNVpA}.
Moreover, experimental triggers in the
fragmentation region of the nucleus, for example a trigger on the number of
knocked-out neutrons, could improve the selection of central events.
This might be particularly important for composite projectiles (small nuclei)
to ensure that all nucleons have hit the large nuclear target.

For ``$pA$'' collisions an analytical solution for the radiation field in the
forward light-cone can be found to first order in the coupling to the proton
field, but to all orders in the density of the nucleus (see below).
In principle, there is a contribution to large multiplicity
events from coupling the radiation field to second (and higher) order to the
proton source, down by additional powers of $\alpha_s$.
That corresponds to attaching additional non-collinear hard gluon lines to
the proton source (collinear emissions are already resummed in
eq.~(\ref{sat_kt2}) through the DGLAP logarithm), see e.g.~\cite{WG}.
As already mentioned above, for the simplest quadratic effective potential
this is in fact the only contribution to large-multiplicity events since
the effective potential exhibits only one classical minimum (i.e.\
the distribution of saturation momenta is a
$\delta$-function). In general, if the potential starts out with some
higher-dimensional operator, the distribution of saturation
momenta is less sharply peaked, or it exhibits several peaks as for the
double well potential above, and so high-multiplicity events can occur
semi-classically. Their contribution dominates when
$\exp(-\Delta S[\rho])>\alpha_s$, where $\Delta S[\rho]$ is the
action relative to that at the global minimum. This
condition determines the ``cut-off'' of the
semi-classical multiplicity distribution.

In light-cone gauge, the classical fields of the sources before the collision,
which occurs at the tip of the light-cone, are given by the single-hadron
(nucleus) solutions
\bea
A^\pm_{1,2} &=& 0~,\nonumber\\
A^i_{1,2} &=& \frac{i}{g}
 U_{1,2}(x_\perp)\, \partial^i\, U^\dagger_{1,2}(x_\perp)~.
\label{single}
\eea
The $U_{1,2}$'s are rotation matrices in color space, 
\be
U_{1,2}(x_t) = \exp\left(-ig \Phi_{1,2}(x_t)\right)~.
\ee
The gauge potentials $\Phi_{1,2}$ satisfy the two-dimensional Poisson equation
\be
-\nabla^2 \Phi_{1,2}(x_t) = g\, \rho_{1,2}(x_t)~.
\ee
The average over the color charge density configurations $\rho_1$ and $\rho_2$
with a Gaussian weight as in eq.~(\ref{aveO}) thus corresponds to an
average over the rotations in color space, $U_{1,2}$.

The fields~(\ref{single})
serve as the initial conditions for solving the Yang-Mills
equations of motion in the forward light cone; they were
solved in~\cite{dm} for the case of asymmetric collisions where 
the classical field of one of the colliding sources is much stronger 
than that of the other source.
Chosing the gauge $x^+ A^- + x^- A^+ =0$, the radiation
field in the forward light cone region  
is given by 
\bea
A^i(\tau,x_t) &=& U(x_t) \left( \beta^i(\tau,x_t) + \frac{i}{g}
\partial^i\right) U^\dagger(x_t)~,\nonumber\\
A^\pm(\tau,x_t) &=& \pm x^\pm \, U(x_t) \, \beta(\tau,x_t) \,
 U^\dagger(x_t)~.  \label{sol_flc}
\eea
Here, $\tau=\sqrt{2x^+ x^-}$ denotes proper time and $x_t$ is the 
transverse coordinate.
The $U$'s are rotation matrices in color space, to be specified shortly.
At asymptotic times, $\tau\to\infty$, the fields $\beta$ and $\beta^i$ are
given by superpositions of plane wave solutions,
\bea
\beta(\tau\to\infty,x_t) &=& \nonumber\\
& &\hspace{-1.5cm} \int\frac{d^2p_t}{(2\pi)^2}
              \frac{1}{\sqrt{2\omega\tau^3}}
          \left\{
           a_1(p_t)e^{ip_t \cdot x_t-i\omega\tau} +c.c.\right\}~,\\
\beta^i(\tau\to\infty,x_t) &=&  \nonumber\\
& &\hspace{-1.5cm} \int\frac{d^2p_t}{(2\pi)^2}
\frac{1}{\sqrt{2\omega\tau}}\frac{\epsilon^{il}p_t^l}{\omega}
          \left\{
          a_2(p_t)e^{ip_t \cdot x_t-i\omega\tau} +c.c.\right\}~.
\eea
The number distribution of produced gluons at rapidity $y$ and transverse
momentum $p_t$ is given by
\be \label{numdis}
\frac{dN}{d^2p_t \, dy} = \frac{2}{(2\pi)^2} {\rm tr}~\left<
\left| a_1(p_t)\right|^2 + \left| a_2(p_t)\right|^2\right>~.
\ee
For $pA$ collisions, the
$U$'s appearing in eq.~(\ref{sol_flc}) are just the $U_2$'s from
eq.~(\ref{single}); that is, to leading order in the weak
field, the plane
waves in the forward light cone are just gauge rotated by the strong field.

The squared amplitudes now have to be averaged over the density configurations.
The radiation number distribution~(\ref{numdis})
depends on the {\em integrated} color charge densities of the sources,
\bea
\chi_p(y) &=& \int\limits_y^{y_p} dy' \, \mu^2_p(y')~,\nonumber\\
\chi_A(y) &=& \int\limits_{y_A}^y dy' \, \mu^2_A(y')~,
\eea
where $y_p=-y_A$ denote the beam rapidities. 
Due to the nuclear enhancement, in the central region
$\chi_A\gg\chi_p$. Up to some proportionality
constant which is of no relevance here, the integrated color charge densities
$\chi_{p,A}$ times $g^4$ 
equal the saturation scales $Q_{p,A}^2$ of the sources.

A high-resolution probe with $p_t\gg Q_{A}(y)$ resolves the gluons in
the target nucleus, i.e.\ it appears dilute at that scale, 
while a low-resolution source with $p_t\lton Q_{A}(y)$ sees a dense,
nearly ``black'' target without structure (see the expressions for the
gluon-nucleus cross section in the various regimes in~\cite{djm1}).
In particular, since $Q_{A}(y)\gg Q_{p}(y)$
at central rapidity, for gluons produced with
transverse momenta $Q_{p}(y)\lton p_t\lton Q_{A}(y)$ the proton projectile
is dilute while the nuclear target is dense. Performing
the integrals over color charge density configurations with a
quadratic potential
for fixed saturation momenta $Q_{A}$ and $Q_{p}$, the transverse momentum
distributions in the two regimes are given 
by~\cite{dm}\footnote{See also the results of ref.~\cite{KM_Kop},
where gluon bremsstrahlung off nuclear targets was obtained by resumming a
Glauber multiple collision series in the dipole model, rather than
solving classical Yang Mills equations.}
\bea
\frac{d N}{d^2b \, d^2p_t \, d y} &\sim&
 \frac{1}{g^2} \frac{Q_A^2(y) \, Q_p^2(y)}{p_t^4}
\log\frac{p_t^2}{Q_A^2(y)} \label{pert_kt4} \\ 
& &\hspace{3.5cm} (p_t> Q_A) \nonumber\\
\frac{d N}{d^2b \, d^2p_t \, d y} &\sim&
\frac{1}{g^2}\, \frac{Q_p^2(y)}
{p_t^2}\log\frac{p_t^2}{Q_p^2(y)}\label{sat_kt2}\\
& &\hspace{3cm} (Q_p\lton p_t\lton Q_A)~. \nonumber
\eea
Again, some proportionality constants which are inessential for the present
discussion have been dropped (see~\cite{dm}).
The logarithm in eq.~(\ref{sat_kt2}) arises from resumming collinear
gluon emissions from the proton source at the scale $p_t^2$~\cite{djm1},
i.e.\ from DGLAP evolution.
For $Q_A\gg Q_p\simeq\Lambda_{\rm QCD}$ the dominant contribution to the
multiplicity density per unit rapidity is from the region~(\ref{sat_kt2}),
\be
\frac{d N}{d^2b\, d y} \sim \frac{1}{g^2}\, Q_p^2(y) \,
\log^2\frac{Q_A^2(y)}{Q_p^2(y)}~.
\ee
The contribution from the high-$p_t$ region~(\ref{pert_kt4}) is
\be
\frac{d N}{d^2b\, d y} \sim \frac{1}{g^2} \, Q_p^2(y)~.
\ee
Thus, fluctuations in the $p_t$-integrated
multiplicity per unit rapidity, $dN/dy$, should be dominated by fluctuations
of the saturation momentum of the {\em proton}. Fluctuations of $Q_A$ do occur
but their effect on $dN/dy$ is logarithmically suppressed, and so the two
quantities are approximately uncorrelated. This can be seen also from the
fact that for a dense nuclear target, the gluon-nucleus~\cite{djm1} and
the quark-nucleus~\cite{DJM2} cross sections are essentially given by
the nuclear area $\pi R_A^2$, which does not vary; once the target
has reached the ``black-body limit'', cross sections are geometrical.
Rather, high-multiplicity events originate from larger than average
density of partons in the dilute projectile proton. Note that fluctuations
of the inelastic $pA$ cross section have been considered previously,
e.g.~ref.~\cite{CT}. What is new here is that the distribution of inelastic
cross sections is related to the distribution of correlation
lengths exhibited by the effective action~(\ref{eff_act}), and that the
transverse momentum distributions of produced particles changes systematically
with multiplicity, see below.

Thus, to first approximation an average over many events within one event
class (which is defined by $dN/dy$) corresponds to an
average\footnote{It is represented here as an integral over a continuous
distribution of curvatures but of course $w(\mu^2)$ could just be a sum over
delta-functions if one has a potential with several discrete minima.}
over $Q_A$ at fixed $Q_p$:
\bea
\frac{d N}{d^2p_t \, d y}(\mu^2_p) &=&
\int d\mu^2_A\, w(\mu^2_A) 
\int {\cal D}\rho_p\, e^{-S[\rho_p,\mu^2_p]} \nonumber\\
& &\times\int {\cal D}\rho_A\, e^{-S[\rho_A,\mu^2_A]}
\frac{2}{(2\pi)^2}\,{\rm tr}\, \left(|a_1|^2+|a_2|^2\right)\nonumber\\
&=& \int {\cal D}\rho_p \, e^{-S[\rho_p,\mu^2_p]}
\int {\cal D}\rho_A \, e^{-S[\rho_A,\overline{\mu^2_A}]}\nonumber\\
& &\quad\times\frac{2}{(2\pi)^2}\,{\rm tr}\, \left(|a_1|^2+|a_2|^2\right)~.
\eea
Note the difference to fluctuations of the multiplicity at high
transverse momentum above some (large) fixed cutoff $p_0$.
The integral of the distribution~(\ref{pert_kt4}) from binary hard scatterings
over $p_t>p_0$ (with $p_0>Q_A$) is equally sensitive to fluctuations of
{\em both} the projectile density as well as the target density~\cite{WG}. 

\begin{figure}[htp]
\centerline{\hbox{\epsfig{figure=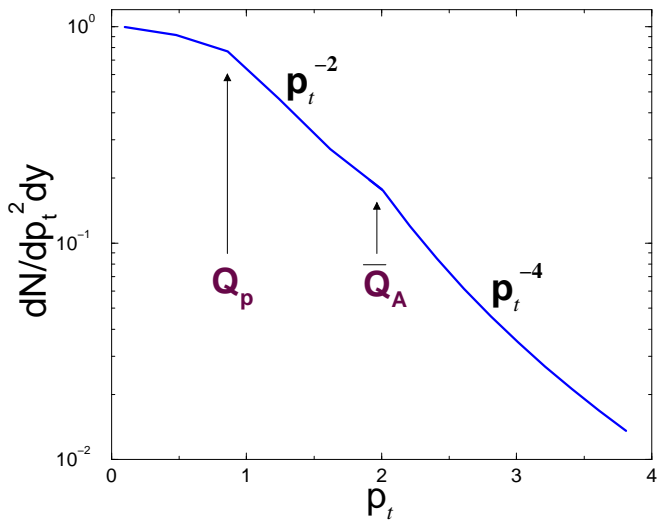,height=7cm}}}
\caption{Schematic distribution of the $p_t$ distribution of produced gluons
in a given class of events, as determined by the saturation momentum of the
proton $Q_p$ (or, equivalently, by $dN/dy$). $\overline{Q}_A$
denotes the average saturation momentum of the nucleus.}
  \label{fig_dNdpt}
\end{figure}
The transverse momentum distribution within one event class therefore
has the shape depicted in Fig.~\ref{fig_dNdpt}. At very small transverse
momentum $p_t\lton Q_p$, the distribution appears rather flat~\cite{KNVpA}.
Here, both the field of the proton as well as that of the nucleus are in the
nonlinear ``saturation'' regime.
As discussed above, the width of this region is proportional to the
square root of
$dN/dy$. For the highest multiplicity classes, when
$dN/dy$ is about $200^{1/3}\simeq 6$ times larger than on average, this
region of transverse momenta could extend all the way up to $\overline{Q}_A$.
In turn, for low multiplicity events it will shrink.

Above $Q_p$, up to about
the average nuclear saturation momentum $\overline{Q}_A$ the gluon distribution
drops approximately like $1/p_t^2$. This is the regime where the proton field
is weak but that of the nucleus still is ``saturated''. It can extend
down to small transverse momentum for the lowest
multiplicity class, or could be ``squeezed'' completely at very large $dN/dy$.

For a closer look at the $p_t$ distributions in different multiplicity
classes it is useful to divide the distributions from each class by that
corresponding to the highest multiplicity bin, forming the ratio
\be
R(p_t;Q_p^2) = \frac{dN}{dp_t^2\, dy}(Q_p^2) \left(
\frac{dN}{dp_t^2\, dy}(Q^2_{p,{\rm max}}) \right)^{-1}~.
\ee
This should be flat at small transverse momenta, $p_t\lton Q_p$. Above $Q_p$
it drops since ${dN}/{dp_t^2\, dy}(Q_p^2)$ decreases while
${dN}/{dp_t^2\, dy}(Q^2_{p,{\rm max}})$ is approximately constant.
Above $Q_A$ finally, the ratio flattens and becomes constant since
from eq.~(\ref{pert_kt4}) the distributions in the various multiplicity bins
become proportional.

\begin{figure}[htp]
\centerline{\hbox{\epsfig{figure=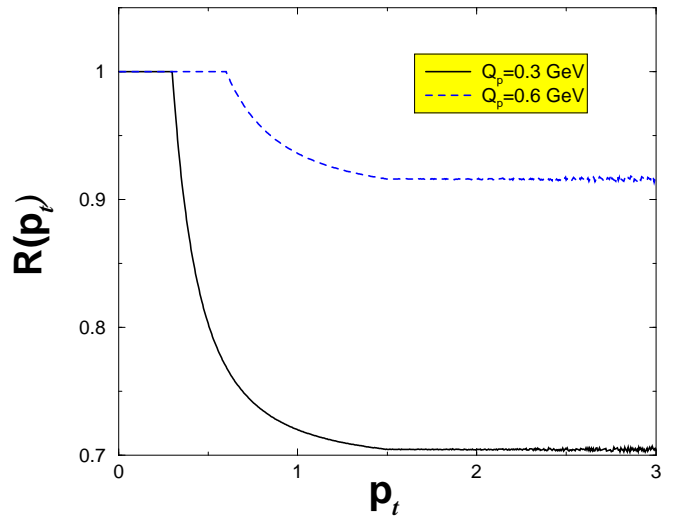,height=7cm}}}
\caption{$p_t$ distributions in various multiplicity bins relative to that
in the highest-multiplicity bin.}
  \label{fig_R}
\end{figure}
Fig.~\ref{fig_R} shows the behavior of $R(p_t)$. For this figure, the
$p_t$ distributions were taken to be
\bea
\frac{dN}{dp_t^2\, dy} &=& \frac{1}{p_t^2+\Lambda^2}~,\quad
   \quad\quad (p_t<Q_p)\\
     &=& \frac{Q_p^2}{p_t^2(Q_p^2+\Lambda^2)}~,\quad (Q_p<p_t<Q_A)\\
     &=& \frac{Q_p^2Q_A^2}{p_t^4(Q_p^2+\Lambda^2)}~,\quad (p_t>Q_A)~,
\eea
with the cutoff $\Lambda=0.2$~GeV, the nuclear saturation scale
$Q_A=1.5$~GeV and $Q_{p,{\rm max}}=Q_A$.
These numbers may be reasonable for the central rapidity region of
$p+Au$ collisions at RHIC energy. For LHC they are expected to be bigger
by about a factor $\exp(\lambda (y_{\rm LHC}-y_{\rm RHIC})/2)\simeq1.7$, where
$y_{\rm LHC}\simeq8.8$ and $y_{\rm RHIC}\simeq 5.4$ are the beam rapidities
at the two colliders, and the intercept $\lambda\simeq 0.3$~\cite{gbw}.

The broadening of the $p_t$ distributions with increasing multiplicity then
leads to an increase of $\langle p_t\rangle$ with $dN/dy$; at least as long as
$Q_p\ll Q_A$, $\langle p_t\rangle$ was shown to scale with the square root of
$dN/dy$~\cite{dm,djm1}. This may lead to changes of multiparticle
correlations~\cite{correl} with increasing $dN/dy$. Moreover,
at midrapidity, the inelastic cross section is
dominated by scattering of small-$x$ gluons and a nearly flavor symmetric
sea of (anti-)quarks. Therefore, when the typical transverse momentum
of produced gluons becomes larger than typical hadronic mass scales, the
hadronic final state should as well be nearly flavor symmetric.
It might be impossible to achieve sufficiently large $\langle p_t\rangle$ at
either RHIC or LHC to be able to neglect, as an example, the masses of the
kaons. Nevertheless, it would be very interesting to see whether
multiplicity ratios for e.g.\ $K^\pm$ to $\pi^\pm$ (or $\eta$ to $\pi^0$)
do increase at all with the multiplicity of the event. Experimentally, this is
not seen for $p\bar{p}$ collisions up to Tevatron
energies~\cite{Alexopoulos:wt}. It might be related to the fact that the
correlation length for thermal Wilson lines {\em at confinement} is large, on
the order of the proton radius~\cite{SDL}.

Finally, it may be interesting to consider the distribution of leading
hadrons (in the proton fragmentation region) which are produced by
fragmentation of large-$x$ quarks from the proton~\cite{DJM2}, similar to
Deep Inelastic $eA$ scattering~\cite{fgs}.
Far from the beam rapidity of the nucleus, its saturation momentum increases
by a factor $\exp(\lambda y)$ relative to that at central rapidity, while in
turn $Q_p$ decreases by the same factor. The quarks from the incident proton
are then scattered to rather large transverse momenta of order $Q_A$, and
the shape of the
forward $p_t$ distribution should depend less on $Q_p$. That is,
$dN/Q_p^2 \,dp_t^2\, dy$ should approach a ``limiting shape'',
which depends rather weakly on the event multiplicity $dN/dy$ (or $Q_p$).

In summary, experimental analysis of the multiplicity distributions in
high-energy $pA$ collisions, and of correlated changes of the $p_t$
distributions could provide some
fundamental insight into the effective action for
small-$x$ gluons, provided that a semi-classical distribution over
saturation momenta exists and that it is not overwhelmed by higher-order
corrections in $\alpha_s$.
\\[1cm]\\
\noindent
{\bf Acknowledgements:} I thank Jamal Jalilian-Marian, Mark Strikman, and
Raju Venugopalan
for useful discussions, the DOE for support from Grant DE-AC02-98CH10886,
and the organizers of the workshop ``Coherent Effects at RHIC and LHC: Initial
Conditions and Hard Probes'', ECT*, Trento (Italy),
October 14-25, 2002, http://pluto.mpi-hd.mpg.de/trento, where part of
this work was presented.


\end{document}